\PassOptionsToPackage{table}{xcolor}
\documentclass[10pt,twocolumn,letterpaper]{article}

\usepackage{cvpr}      
\usepackage[accsupp]{axessibility}

\definecolor{best1}{RGB}{255, 204, 204} 
\definecolor{best2}{RGB}{255, 230, 204} 
\definecolor{best3}{RGB}{255, 255, 204} 

\definecolor{cvprblue}{rgb}{0.21,0.49,0.74}
\usepackage[pagebackref,breaklinks,colorlinks,allcolors=cvprblue]{hyperref}

\def\paperID{2450} 
\def\confName{CVPR}
\def\confYear{2025}

\title{BG-Triangle: Bézier Gaussian Triangle for 3D Vectorization and Rendering}

\author{
Minye Wu$^{1,*}$
\qquad
Haizhao Dai$^{2, 3,*}$
\qquad
Kaixin Yao$^{2}$ 
\qquad
Tinne Tuytelaars$^{1,\dagger}$
\qquad
Jingyi Yu$^{2,\dagger}$
\\
$^{1}$ KU Leuven \qquad $^{2}$ ShanghaiTech University \qquad $^{3}$ Cellverse Co, Ltd. \\
}

\begin{document}

\maketitle
{\let\thefootnote\relax\footnote{* Authors contributed equally.
${\dagger}$ The corresponding authors.
}}
\begin{abstract}

Differentiable rendering enables efficient optimization by allowing gradients to be computed through the rendering process, facilitating 3D reconstruction, inverse rendering and neural scene representation learning. To ensure differentiability, existing solutions approximate or re-formulate traditional rendering operations using smooth, probabilistic proxies such as volumes or Gaussian primitives. Consequently, they struggle to preserve sharp edges due to the lack of explicit boundary definitions. We present a novel hybrid representation, Bézier Gaussian Triangle (BG-Triangle), that combines Bézier triangle-based vector graphics primitives with Gaussian-based probabilistic models, to maintain accurate shape modeling while conducting resolution-independent differentiable rendering. We present a robust and effective discontinuity-aware rendering technique to reduce uncertainties at object boundaries. We also employ an adaptive densification and pruning scheme for efficient training while reliably handling level-of-detail (LoD) variations. Experiments show that BG-Triangle achieves comparable rendering quality as 3DGS \cite{kerbl3Dgaussians} but with superior boundary preservation. More importantly, BG-Triangle uses a much smaller number of primitives than its alternatives, showcasing the benefits of vectorized graphics primitives and the potential to bridge the gap between classic and emerging representations.

\end{abstract}
    
\section{Introduction}
\label{sec:intro}

Scene representations lie at the heart of computer graphics and 3D computer vision. Over time, they have undergone significant evolution and led to profound changes in modeling and rendering techniques as well as hardware architectures. Traditional representations such as meshes, point clouds, and volumes are fundamental in defining how 3D structures are reconstructed (e.g., as point clouds from multi-view images~\cite{hartley2003multiple, furukawa2009accurate, snavely2006photo} or as volumes from CT reconstruction~\cite{shepp1974fourier}) and visualized (e.g., via rasterization~\cite{foley1996computer}, ray tracing~\cite{whitted2005improved} or volume rendering~\cite{drebin1988volume}). In the era of deep learning, differentiable rendering, by enabling gradients to flow through the rendering process, has introduced a significant shift in 3D representations. At its core, differential rendering allows optimization algorithms to directly update the 3D representation based on a rendered image's error (e.g., differences from a target image), making it possible to integrate these representations into end-to-end learning frameworks.

\begin{figure}[t]
    \centering
    \includegraphics[width=\linewidth]{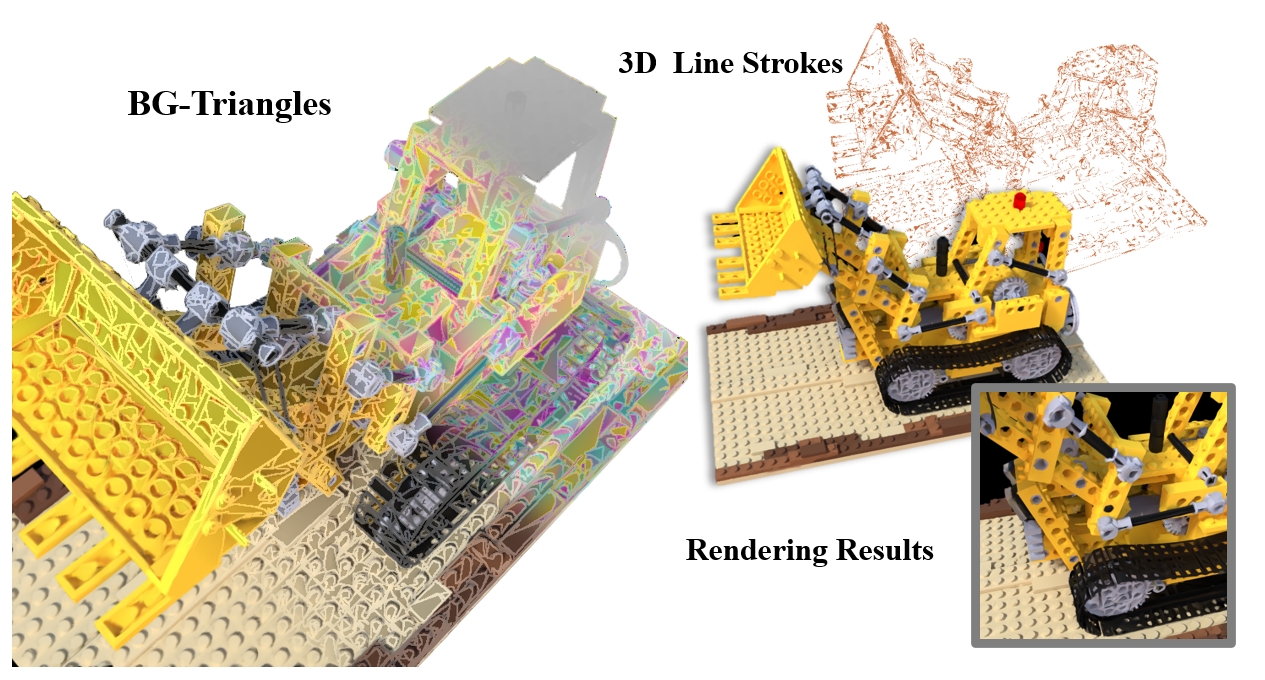}
    \caption{BG-Triangle can transform multi-view images into vectorized 3D scenes, enabling the rendering of sharp  novel views.}
    \label{fig:teaser}
\end{figure}

Neural Radiance Fields or NeRFs~\cite{mildenhall2020nerf}, for example, leverage neural networks to encode radiance and density fields in a volumetric space to overcome resolution limitations and enable view dependency in classic volume rendering. Using GPU-specific data structures and acceleration schemes, InstantNGP~\cite{muller2022instant} stores and optimizes features in local space to achieve extremely fast training and rendering. 3D Gaussian Splatting (3DGS)~\cite{kerbl3Dgaussians}, on the other hand, revives point-based rendering by adopting Gaussian primitives to give the point cloud a certain volume. It combines the flexibility of explicit point clouds with the powerful expressive capability of volumetric representation for complex scenes. 3DGS can be directly implemented using the standard rasterization pipeline, including classical rendering tricks such as mip-mapping~\cite{williams1983pyramidal} for Level-of-Detail (LoD) controls~\cite{yu2024mip}. Coupled with Structure-from-Motion, both NeRF and 3DGS can be deployed to capture real scenes while preserving correct lighting, reflections, and occlusions challenging for traditional techniques. 

A major challenge in 3DGS and NeRF though is that they struggle to preserve sharp edges, especially in close-up views. In 3DGS, the use of Gaussian kernels inherently blurs detail because each kernel represents a region in 3D space with a soft, rounded distribution. This limits the model's ability to capture sharp transitions, as the Gaussians overlap and blend into one another. Similarly, NeRFs smooth out rapid transitions due to the continuous nature of the recovered density fields. Edge-aware regularization or tailored sampling schemes can potentially mitigate the problem, but the fundamental issue lies in the representation. If we plot the representation spectrum from deterministic to probabilistic, at one end are 3D meshes with explicit connectivity to preserve shape but not readily differentiable. At the other end are volumes/NeRFs that correspond to probabilistic models with a well-defined gradient to facilitate optimization but lack explicit shape controls. 3DGS lies somewhere in between: it adjusts the shape of the probabilistic model to reflect spatial discontinuities but still insufficient to model sharp change such as edges. In fact, uncertainties in 3D space increase when we render viewpoints far away from the training perspectives, e.g., close-up views. Consequently, both NeRF and 3DGS produce visible blurs especially near occlusion boundaries.

To address this challenge, we set out to locate, on the representation spectrum, a new representation that has a comparable level of certainty as explicitly defined meshes while maintaining flexibility and differentiability as the probabilistic models. We present Bézier Gaussian Triangle or BG-Triangle as shown in Fig. \ref{fig:teaser}, a hybrid representation that uses vector graphics as shape descriptors and Gaussians as pixel-level primitives. At the shape-level, we exploit Bézier surfaces as a 3D scene representation, where each primitive parameterizes a local surface region of the scene defined by a set of control points and attributes. These parametric surfaces can be tessellated into triangular meshes to support the rasterization pipeline, generate attribute buffers, perform a depth test, etc. At the pixel-level, we generate pixel-aligned Gaussian primitives based on the attribute buffers as a rendering proxy, thereby achieving resolution-independent rendering. This further allows for visual compensation via alpha blending during rendering and facilitates smooth gradient computation during optimization. To render a BG-Triangle, we adopt a discontinuity-aware rendering scheme that assigns a blending coefficient to each pixel of the Gaussian primitive after conducting splatting, to reflect the influence of the BG-Triangle primitive to which it belongs. This mechanism effectively suppresses the probabilistic uncertainties outside the primitive regions. We further develop a tile-based, boundary-aware rendering technique to achieve real-time performance.

For training, we introduce a splitting and pruning process that allows the reconstruction from only a coarse point cloud initialization. This process resembles those used on adaptive subdivision surfaces~\cite{amresh2003adaptive}, which subdivide surfaces that lack sufficient representational capabilities in the local space. In a similar vein, BG-Triangle, once trained, can represent objects with varying levels of granularities. At coarser levels, a primitive shares the same set of attributes to represent large areas, making it highly efficient in utilizing information. This efficiency allows for high quality rendering even with a very small number of primitives. 

Comprehensive experiments demonstrate that, BG-Triangle, with vectorized primitives, can effectively preserve sharp boundaries without compromising the overall rendering quality. In particular, it supports a viewer to zoom into fine geometric details that are difficult to achieve using existing techniques. With a compact set of parameters, BG-Triangle strikes an new balance between parameter efficiency and rendering quality/sharpness. Compared with 3DGS, BG-Triangle lies on the representation spectrum one step closer to the deterministic end and may stimulate new representations in the near future.

\section{Related Work}

\begin{figure*}[ht]
    \centering
    \includegraphics[width=0.95\linewidth]{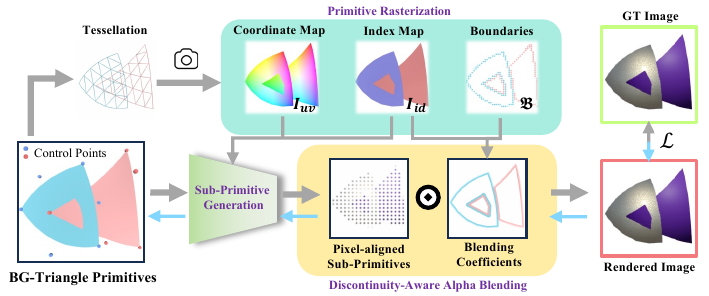}
    \caption{\textbf{Rendering pipeline of BG-Triangles.} The BG-Triangle rendering pipeline has three modules. The Primitive Rasterization module tessellates the Bézier triangle, producing coordinate and index maps, as well as boundary points. These maps are then used in the sub-primitive generation module to create pixel-aligned sub-primitives for differentiable rendering. The discontinuity-aware alpha blending module utilizes the boundary points to render images with sharp edges. Finally, the fully differentiable pipeline allows gradient backpropagation along the blue arrows to optimize the control points of BG-Triangles.}
    \label{fig:pipeline}
    \vspace{-2mm}
\end{figure*}

\paragraph{Traditional Scene Modeling and Rendering.}

Multi-view reconstruction has long been a fundamental problem in computer vision. Structure from Motion \cite{snavely2006photo, Schonberger_2016_CVPR} focuses on generating point cloud scene representations during camera calibration, which is a basic approach for scene modeling. Some methods \cite{kazhdan2006poisson,hoppe1992surface,bernardini1999ball} can reconstruct meshes based on point clouds, but often suffer from issues such as topological errors, holes, and challenges in handling boundaries, which affect the completeness and accuracy of the meshes. Multi-View Stereo \cite{MVS} methods have evolved to produce dense point clouds and detailed meshes for high-fidelity 3D reconstructions. Once high-quality meshes are obtained, various rendering techniques can be used to enhance visual realism, including neural texture blending \cite{buehler2001unstructured, xiang2021neutex, thies2019deferred}, image-based rendering \cite{debevec1996modeling,shum2008image, wang2021ibrnet}, and deep surface light fields \cite{chen2018deep}. However, the effectiveness of these rendering methods heavily depends on the quality of the underlying mesh, while representing objects with smooth surfaces or complex geometries remains challenging.

Mesh-based differentiable rendering (DR) methods enable end-to-end optimization for meshes. Some of them exploit analytical derivative of rendering~\cite{liu2018paparazzi, azinovic2019inverse} or  approximated gradients~\cite{loper2014opendr, genova2018unsupervised, liu2019soft, pidhorskyi2024rasterized}. However, they rely heavily on mesh initialization, leading to slow convergence and incomplete reconstructions with poor initialization. Our method, using a coarse point cloud, offers greater flexibility than traditional DR. While recent techniques~\cite{shen2023flexible, munkberg2022extracting} improve geometry reconstruction but use discrete texture maps, we vectorize texture boundaries.

\paragraph{Neural Scene Representation.}



In recent years, Neural Radiance Fields (NeRF) \cite{mildenhall2020nerf} have achieved remarkable progress in 3D scene reconstruction by using neural networks as implicit representations to jointly model both geometry and appearance, which enables photo-realistic rendering with fine details in lighting \cite{nerv2021,zhang2021nerfactor} and materials \cite{barron2021mip,barron2022mip}. However, the implicit nature of these representations makes scene editing a major challenge. Although some approaches have been developed to introduce editing capabilities \cite{liu2021editing,zhang2021stnerf}, they remain computationally intensive. 
Some methods~\cite{chen2022mobilenerf, tang2022nerf2mesh} embed radiance fields into precomputed meshes for rasterization but suffer from geometric inaccuracies and limited texture resolution. 
Other methods \cite{wang2021neus,instant-nsr} attempt to extract explicit geometry from implicit representations using signed distance functions.
%
They treat objects as a whole to effectively reconstruct geometry. In contrast, our approach describes scenes differently by vectorizing the 3D structure of geometry and texture boundaries. 


In contrast, 3D Gaussian Splatting (3DGS) \cite{kerbl3Dgaussians} uses an explicit representation that models scenes through a set of Gaussians. This flexible primitive allows 3DGS to effectively capture intricate geometric and appearance details. However, this high degree of flexibility can also lead to disorganized scene structures. Scaffold GS \cite{scaffoldgs} tackles this issue by using anchor-based points to guide the Gaussian arrangement, though the constraints between Gaussians and anchors remain limited. BG-Triangle, on the other hand, imposes hard constraints to regulate the local regions within each primitive.


\paragraph{Vector Graphics.}

Vector graphics have been a fundamental part in computer graphics for decades, they are used to represent stylized images, such as line drawings \cite{egiazarian2020deep,favreau2016fidelity,noris2013topology}, fonts \cite{lopes2019learned}, clip-art \cite{favreau2017photo2clipart,dominici2020polyfit},
image triangulations \cite{lawonn2019stylized}, and sketches \cite{su2021marvel,bhunia2021vectorization}. The vector primitives provide the key advantages of infinite scalability without quality loss, flexible edit ability, and efficient rendering. 

Currently, neural networks have also impacted vector graphics. DiffVG \cite{Li:2020:DVG} leverages differentiable rasterization and uses gradient descent controlled by Bézier curves to generate vector graphics. Im2Vec \cite{reddy2021im2vec} utilizes variational auto encoders for image vectorization, mapping raster images into a latent space and then decoding them into Bézier curves. These methods are limited to 2D shape fitting and can only approximate basic shapes and colors, making them unsuitable for direct application in 3D scenarios. Some methods \cite{L_pez_2020, requicha1982solid} apply vector graphics in the 3D domain. However, they require pre-existing 3D geometry, such as a mesh, as input. In contrast, our method directly reconstructs 3D vector primitives from multi-view images.



\section{Bézier Gaussian Triangles}

We first introduce our basic primitive, Bézier Gaussian Triangle in, Sec.~\ref{sec:bezier-gaussian-triangle}. Then, we discuss the design of our rendering pipeline in Sec.~\ref{sec:rendering} to achieve efficient differentiable rendering, along with the optimization strategies in Sec.~\ref{sec:Optimization}.

\subsection{Bézier Patches }
\label{sec:bezier-gaussian-triangle}

Bézier triangle surfaces \cite{farin1986triangular} are a type of parametric surface that extends Bézier curves into three dimensions, defined over a triangular domain. It efficiently represents complex surfaces and curves, reducing the need for excessive triangular facets as in traditional mesh representations, and has a well-defined barycentric coordinate system.

Specifically, a Bézier triangle \(\mathbf{S}\) of degree \(n\) is defined by its control points, denoted as \(\mathbf{p}_{i, j, k} \in \mathbb{R}^{3}\), where \(i + j + k = n\) and \(i, j, k \ge 0\).
Using the barycentric coordinates \((u, v, w)\), where \(u + v + w = 1\) and \(u, v, w \ge 0\), a point \(\mathbf{S}(u, v, w) \in \mathbb{R}^{3}\) on the Bézier triangle can be uniquely determined by the barycentric interpolation formula:
\begin{equation}\label{eq:barycentric-interpolation}
    \mathbf{S}(u, v, w) = \sum_{i = 0}^{n} \sum_{j = 0}^{n - i} B_{i, j, k}^{n}(u, v, w) \mathbf{p}_{i, j, k},
\end{equation}
where:
\begin{equation}
    B_{i, j, k}^{n}(u, v, w) = \frac{n!}{i! j! k!} u^{i} v^{j} w^{k},
\end{equation}
is the Bernstein polynomial of degree \(n\) defined on the barycentric coordinate system. Based on the Bézier triangle surface, we propose Bézier Gaussian triangle (BG-Triangle) as the representation primitive. 

\subsection{Rendering Pipeline}\label{sec:rendering}
A major advantage of BG-Triangle is that it directly supports differentiable rendering while maintaining sharp boundaries. Fig.~\ref{fig:pipeline} shows our rendering pipeline. 

\vspace{2mm}
\textbf{Tessellation and Rasterization.} 
The mapping between screen pixels and a Bézier triangle is not straightforward due to the inherent non-linearity and parametric nature of its formulation. To align with the pixel grid in the image plane, we need to tessellate the Bézier triangle into smaller flat triangles that approximate the curved surface. The vertices of these flat triangles inherit attribute information from the original primitive, including barycentric coordinates and the primitive's ID. Using a traditional rasterization pipeline, we render a raster image after depth testing and interpolate these attributes across the triangles to generate a coordinates map $\mathbf{I}_{uv} \in \mathbb{R}^{H \times W \times 3}$ and an index map $\mathbf{I}_{id} \in \mathbb{Z}^{H \times W}$, where $H$ and $W$ are the height and width of the image. These attribute buffers indicate the coordinates and primitive ID for each pixel in the target view. Pixels with neighboring pixels assigned different primitive IDs are marked as boundary pixels and stored in a 2D boundary point set $\mathcal{B}$ for use in discontinuity-aware rendering.

\begin{figure}[t]
    \centering
    \includegraphics[width=0.95\linewidth]{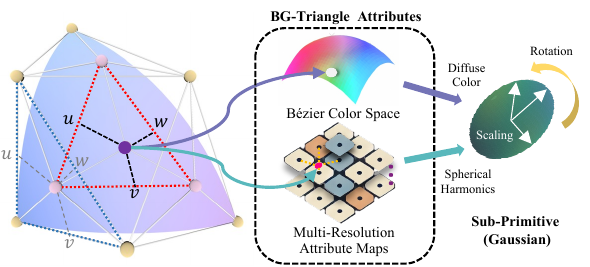}
    \caption{\textbf{Sub-Primitive Generation.} 
    For a differentiable rendering pipeline, we generate 3D Gaussian sub-primitives on a degree-2 Bézier triangle (left). Coordinates  (purple) are computed from control points (yellow) via barycentric interpolation, while other attributes are interpolated barycentrically or via a multi-resolution attribute map.}
    \label{fig:sub_gen}
\end{figure}

\vspace{2mm}
\textbf{Sub-Primitive Generation.}  
\label{sub-primitive}
After tessellation and rasterization, the primitives form the layout structure and contours of the scene. However, this is insufficient to describe the appearance. Also, the rendering pipeline is not differentiable. Therefore, we propose using a Gaussian-based probabilistic model, inspired by 3DGS, to generate pixel-aligned sub-primitives on the image pixels. Note that, unlike in 3DGS, the Gaussians in our setting are generated on the fly, i.e. a different viewpoint will be rendered using a different set of Gaussians, sampled from the same BG Triangle.

We first compute the 3D coordinates for each foreground pixel $\mathbf{q}$ in the image using Equ.~\eqref{eq:barycentric-interpolation}. Specifically, we retrieve the BG-Triangle primitive index $\mathbf{I}_{id}(\mathbf{q})$ and its barycentric coordinates $\mathbf{I}_{uv}(\mathbf{q})$. The 3D coordinate $\mathbf{S}_{\mathbf{q}}$ on the BG-Triangle corresponding to pixel $\mathbf{q}$ on the image is given by:
\begin{equation}\label{equ:coordinate-interpolation}
    \mathbf{S}_{\mathbf{q}}(\mathbf{I}_{uv}(\mathbf{q}), \mathbf{I}_{id}(\mathbf{q})) = \sum_{i = 0}^{n} \sum_{j = 0}^{n - i} B_{i, j, k}^{n}(\mathbf{I}_{uv}(\mathbf{q})) \mathbf{p}_{i, j, k}(\mathbf{I}_{id}(\mathbf{q})),
\end{equation}
where $\mathbf{p}_{i, j, k}(\mathbf{I}_{id}(\mathbf{q}))$ is a control point of the BG-Triangle primitive indexed by $\mathbf{I}_{id}(\mathbf{q})$.

Then, at each of these coordinate points $\{\mathbf{S}_{\mathbf{q}}\}$, we place Gaussians that, similar to 3DGS, possess attributes including rotation, scaling, and spherical harmonic (SH) coefficients, but not opacity, since the geometry is defined by the Bézier triangles. The attributes of these sub-primitives are interpolated based on the color control points $\{\mathbf{p}_{i, j, k}^{c}\}$ and attribute map $\mathbf{M}_{h}$ of the BG-Triangle. We assume that the diffuse colors of Gaussians associated with the same BG-Triangle are highly similar and transition smoothly, which aids in gradient computation and optimization of the control points. To enforce this property, we exploit techniques similar to Equ.~\eqref{equ:coordinate-interpolation}. Each color control point $\mathbf{p}_{i, j, k}^c \in \mathbb{R}^{3}$ represents the RGB color channels. We use the following formulation to interpolate the diffuse color $\mathbf{c}_{\mathbf{q}}$:
\begin{equation}
    \mathbf{c}_{\mathbf{q}}(\mathbf{I}_{uv}(\mathbf{q}), \mathbf{I}_{id}(\mathbf{q})) = \sum_{i = 0}^{n} \sum_{j = 0}^{n - i} B_{i, j, k}^{n}(\mathbf{I}_{uv}(\mathbf{q})) \mathbf{p}_{i, j, k}^{c}(\mathbf{I}_{id}(\mathbf{q})),
\end{equation}
For the remaining attributes, we interpolate them using a multi-resolution 2D attribute map based on the pixel's barycentric coordinates. This is formulated as:
\begin{equation}\label{equ:attribute-interpolation}
    \mathbf{a}_h(\mathbf{q}) = \Theta(\mathbf{M}_h(\mathbf{I}_{id}(\mathbf{q})), \mathbf{I}_{uv}(\mathbf{q})),
\end{equation}
where $\Theta(\cdot, \cdot)$ denotes the 2D texture interpolation function, and $\mathbf{a}_h(\mathbf{q})$ represents one attribute for the Gaussian. Here, $h \in \{\text{rotation, scaling, SH coefficients}\}$ refers to the respective Gaussian attributes. These attributes have varying sensitivity to rendering quality, so we apply different resolutions to their attribute maps $\mathbf{M}_h$. The complete generation process is shown in Fig. \ref{fig:sub_gen}.

\vspace{2mm}
\textbf{Discontinuity-Aware Alpha Blending.} 
We use 3DGS \cite{kerbl3Dgaussians} to render these sub-primitives. Since our sub-primitives are Gaussian-based, its uncertainty can introduce a blurring effect during rendering. Within each BG-Triangle, our assumptions and interpolation methods ensure a smooth transition, so any internal blurring does not compromise the visual quality. However, in boundary regions, a mechanism is needed to mitigate this uncertainty. Directly truncating 3D Gaussians at the boundaries results in non-differentiable rendering. We propose to soften the boundary and apply blending coefficients on splatted sub-primitives near boundaries to suppress the uncertainty.

First, we determine the area of influence for the softened boundaries on the image. Given the 2D boundary points $\mathcal{B}$, $\mathbf{I}_{id}$, and $\mathbf{I}_{uv}$, we use Equ.~\eqref{equ:coordinate-interpolation} to infer the corresponding 3D boundary points. Each boundary point is treated as an isotropic Gaussian with a fixed scaling factor $r_b$. Next, we compute the boundary radius $\sigma_i$ of these boundary Gaussians when projected onto the image, which defines the influence range for the $i$-th boundary point $\mathbf{b}_{i} \in \mathcal{B}$. 
Then, we splat the Gaussian-based sub-primitives onto the image and determine the blending coefficient $w(\mathbf{q})$ for each pixel $\mathbf{q}$ covered by the splatted Gaussian $\mathcal{G}$. The blending coefficient is defined as:
\begin{equation} \label{equ:blending_weight}
    w(\mathbf{q}) = \begin{cases}
    0, & \text{if } \mathbf{I}_{id}(\mathbf{b}_{i})<> g, \\
    \gamma(\|\mathbf{q} - \mathbf{b}_{i}\|_{2}; \sigma_i) & \text{if }  \mathbf{I}_{id}(\mathbf{b}_{i}) = \mathbf{I}_{id}(\mathbf{q}), \\
    1 - \gamma(\|\mathbf{q} - \mathbf{b}_{i}\|_{2}; \sigma_i) & \text{otherwise},
    \end{cases}
\end{equation}
where \(g\) is the identifier of the primitive to which this Gaussian $\mathcal{G}$ belongs, and $\gamma$ is a blurring function defined as:
\begin{equation}
    \label{eq:gamma}
    \gamma(d; \sigma) = \min\left(2^{\frac{d}{\sigma} - 1}, 1\right),
\end{equation}
which maps the distance $d$ to $[0.5, 1.0]$ within the boundary radius $\sigma$, ensuring Equ. \eqref{equ:blending_weight} transitions the blending coefficient from $1.0$ to $0.0$ within the boundary region while remaining well defined outside.

\begin{figure}[t]
    \centering
    \includegraphics[width=\linewidth]{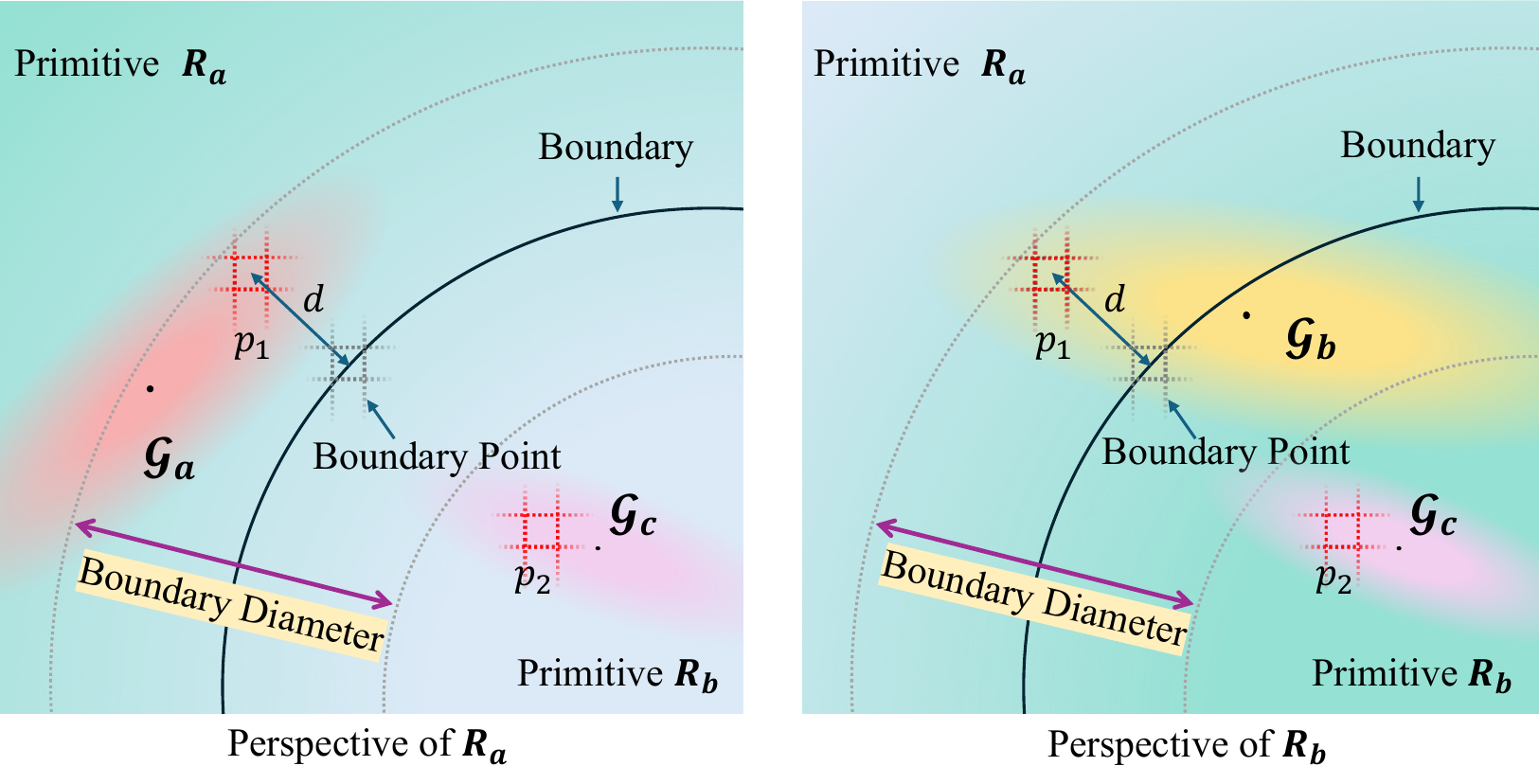}
    \vspace{-6mm}
    \caption{\textbf{Discontinuity-Aware Alpha Blending.} Primitives \(\mathbf{R}_{a}\) and \(\mathbf{R}_{b}\) share a boundary. For pixel \(p_{1}\), which lies within both the boundary region and \(\mathbf{R}_{a}\), the blending coefficient from Gaussian \(\mathcal{G}_{a}\) (associated with \(\mathbf{R}_{a}\)) is larger than that from \(\mathcal{G}_{b}\) (associated with \(\mathbf{R}_{b}\)). In contrast, pixel \(p_{2}\), located well inside \(\mathbf{R}_{b}\) and away from the boundary, is unaffected by Gaussians \(\mathcal{G}_{a}\) and \(\mathcal{G}_{c}\) (not associated with \(\mathbf{R}_{a}\) nor \(\mathbf{R}_{b}\)).}
    \label{fig:blur-weight}
     \vspace{-2mm}
\end{figure}

Finally, to soften the influence of boundaries on the Gaussian \(\mathcal{G}\) at pixel \(\mathbf{q}\), the alpha value \(\alpha(\mathbf{q})\), blended with the contribution \(w(\mathbf{q})\) is given by:
\begin{equation}
    \alpha(\mathbf{q}) = o \cdot w(\mathbf{q}) \cdot e^{-\frac{1}{2}(\mathbf{q} - \mathbf{\mu})^{\top} \Sigma^{-1} (\mathbf{q} - \mathbf{\mu})},
\end{equation}
where $o$ is a constant number close to $1.0$ to help the optimization, \(\mu\) and \(\Sigma\) are the 2D position and 2D covariance matrix projected from \(\mathcal{G}\). Then a point-based alpha blending~\cite{kopanas2022neural} is applied to obtain the final color of the pixel. In Fig. \ref{fig:blur-weight}, we provide examples for better understanding.

\textbf{Rendering Acceleration.} 
Computing blending coefficient $w(\mathbf{q})$ has a very high computational complexity, as it requires iterating through each boundary point in $\mathcal{B}$ for every sub-primitive and each pixel to compute the result. Therefore, when calculating $w(\mathbf{q})$ , we use a tile-based rendering~\cite{fuchs1989pixel} algorithm. This algorithm divides the image space into multiple tiles and pre-computes the range of tiles influenced by each boundary point. We then restrict our search to the relevant tiles to identify the boundary point with the greatest influence, i.e., the one that produces the smallest $\gamma$ value.
To quickly locate boundary points belonging to a specific primitive ID, we sort the boundary points within each tile by ID. This allows us to use binary search to efficiently find the target points. Please refer to the supplementary materials for implementation details of the algorithm.

\subsection{Training and Optimization}
\label{sec:Optimization}
Similar to 3DGS, BG-Triangle directly supports differentiable rendering. Specifically, we apply photometric loss on rendered images and use a loss function that combines the $\mathcal{L}_{2}$ loss with a D-SSIM term as:
\begin{equation}
    \mathcal{L} = (1 - \lambda)\mathcal{L}_{2} + \lambda\mathcal{L}_{\text{D-SSIM}}
\end{equation}
where we set $\lambda = 0.2$ for all our experiments.

\vspace{2mm}
\textbf{Backward Derivative.} 
In Sec.~\ref{sec:rendering}, we use Equ.~\eqref{eq:barycentric-interpolation} to calculate the 3D positions of the sub-primitives and boundary points. This establishes a connection between the control points of the BG-Triangle and these 3D points. Moreover, Equ.~\eqref{equ:blending_weight} is differentiable. Thus, by applying the chain rule, we can compute the derivatives of the control points.  Please refer to the supplementary materials for more details.

\vspace{2mm}
\textbf{Splitting and Pruning.} 
There are many situations that can make optimization challenging, such as overlapping primitives or using excessively large primitives to fit a fine structure. To this end, we propose a splitting and pruning scheme tailored for Bézier Gaussian triangles.

We select primitives for splitting with two criteria: The  gradient amplitude of the control points and edge priors in training views.  A large norm of the average gradient for a set of control points indicates that the primitive requires significant adjustment, often due to under-reconstructed geometry. Thus, we split primitives with an average gradient above $\tau_g$ to better fit finer details. In terms of edge priors, we apply edge detection to the training images to extract edge gradients and back-project these gradient values onto the corresponding primitives. If the accumulated edge gradients on a primitive exceed a certain threshold $\tau_b$, the primitive is split.




Pruning ineffective primitives can improve training efficiency. We use three criteria to identify the primitives that should be pruned: visibility, area, and aspect ratio of the primitives. Here, visibility represents the percentage of training views in which a primitive is visible. This can filter out primitives that are occluded.




\section{Experiments}

We first validate our BG-Triangle representation on toy datasets (Sec.~\ref{sec:proof-of-concept}) and then evaluate its performance on the NeRF Synthetic \cite{mildenhall2020nerf} and masked Tanks \& Temples \cite{knapitsch2017tanks} datasets. We follow the same evaluation protocol for all datasets. For more implementation details, please refer to the supplementary materials.

\begin{figure}[t]
    \centering
    \includegraphics[width=\linewidth]{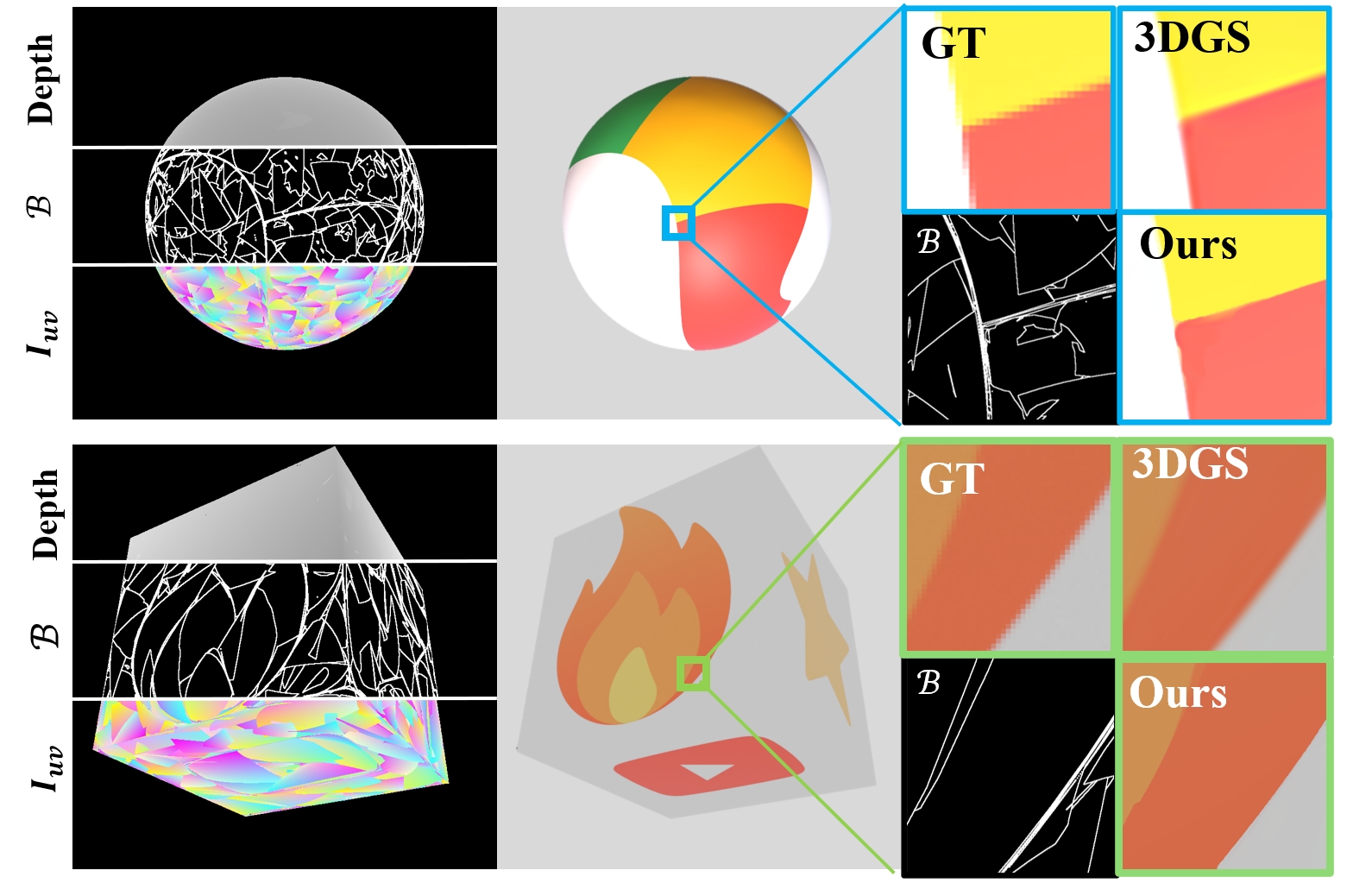}
    \caption{\textbf{Initial validation.} Our method is capable of fitting the object's surface, silhouette, and boundaries by using Bézier triangles, maintaining sharp edges even when zoomed in, and sometimes appearing clearer than the ground truth. In contrast, 3DGS shows blurring under magnification.}
    \label{fig:POC}
\end{figure}

\subsection{Initial Validations}
\label{sec:proof-of-concept}

We render two sets of synthetic data to validate the effectiveness of our method. These datasets consist of a cube and a ball, each with a textured pattern on their geometric surfaces. Cameras are uniformly placed on a sphere, facing the center of the object. Each scene has 100 training views.

As shown in the Fig.~\ref{fig:POC}, our method reconstructs and renders objects from an initial sparse point cloud. 
After training, Bézier primitives effectively outline the object’s silhouette, capturing smooth curves like those in an ellipsoid. The texture is represented by the 3D boundaries, creating recognizable patterns in the boundary map. This structure enables discontinuity-aware rendering, producing sharp, defined edges and superior clarity. In contrast, 3DGS, which aligns only with the pixel sampling rate from training viewpoints, results in blurry close-ups and requires more parameters.

\subsection{Comparisons}

\begin{figure*}[ht]
    \centering
    \includegraphics[width=\linewidth]{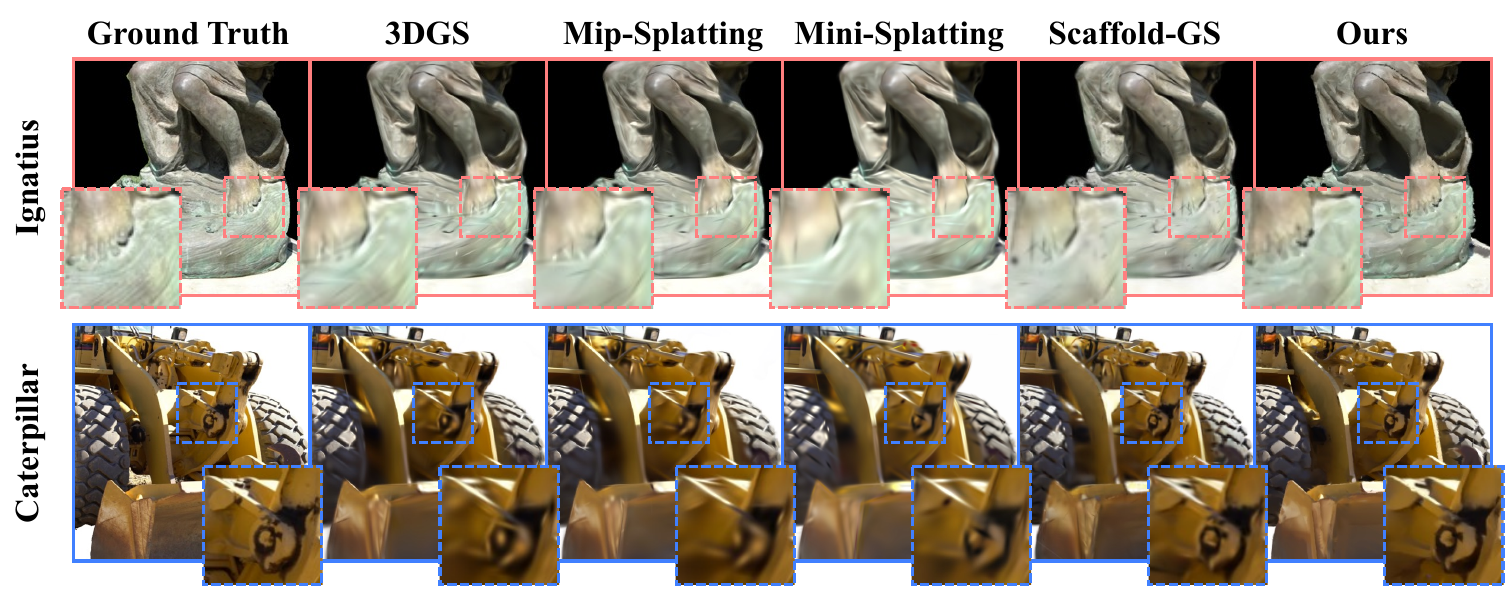}
    \caption{\textbf{Qualitative Results on the Masked Tank \& Temples dataset.} We compare methods with similar  parameter counts. 
These images are rendered with a closer camera perspective. The original images are cropped and utilized as the pseudo ground truth.
    }
    \label{fig:comparison-tt}
\end{figure*}

\begin{table*}[]
\centering
\small
\colorbox{best1}{best} \colorbox{best2}{second-best}  \colorbox{best3}{third-best}

\begin{tabular}{c|cccc|cccc}
\hline
Dataset         & \multicolumn{4}{c|}{NeRF Synthetic}                                                                                  & \multicolumn{4}{c}{Tanks \& Temple}                                                                                 \\ \hline
Method | Metric & \multicolumn{1}{l}{\#Parameters} & \multicolumn{1}{l}{SSIM↑} & \multicolumn{1}{l}{PSNR↑} & \multicolumn{1}{l|}{LPIPS↓} & \multicolumn{1}{l}{\#Parameters} & \multicolumn{1}{l}{SSIM↑} & \multicolumn{1}{l}{PSNR↑} & \multicolumn{1}{l}{LPIPS↓} \\ \hline
3DGS (Full)                     & 17.07M                         & 0.969                     & 33.79                     & 0.031                       & 19.28M                         & 0.959                     & 29.42                     & 0.049                      \\
Mip-Splatting (Full)           & 16.79M                         & 0.970                     & 33.92                     & 0.029                       & 23.37M                         & 0.965                     & 30.04                     & 0.043                      \\
Mini-Splatting (Full)      & 7.274M                         & 0.964                     & 32.39                     & 0.034                       & 7.063M                         & 0.956                     & 28.35                     & 0.053                      \\
Scafold-GS (Full)             & 2.335M                          & 0.965                     & 33.08                     & 0.037                       & 12.82M                         & 0.954                     & 28.47                     & 0.055                      \\ \hline
3DGS                     & 383.5K                         & 0.922                     & 27.18                     & 0.103                       & 855.5K                         & \cellcolor{best1}0.931                     & \cellcolor{best1}27.56                     & \cellcolor{best2}0.099                      \\
Mip-Splatting            & 383.5K                         & 0.919                     & 27.01                     & 0.107                       & 855.5K                         & 0.923                     & \cellcolor{best3}27.20                     & 0.122                      \\
Mini-Splatting           & 339.8K                         & \cellcolor{best3}0.928                     & \cellcolor{best2}28.45                     & \cellcolor{best3}0.093                       & 721.0K                         & 0.919                     & 26.74                     & 0.114                      \\
Scafold-GS               & 360.1K                         & \cellcolor{best2}0.935                     & \cellcolor{best3}28.36                     & \cellcolor{best2}0.081                       & 888.4K                         & \cellcolor{best3}0.924                     & 26.90                     & \cellcolor{best3}0.104                      \\ \hline
\textbf{Ours}                     & 343.5K                         & \cellcolor{best1}0.937                     & \cellcolor{best1} 29.16                     & \cellcolor{best1}0.050                       & 867.1K                         & \cellcolor{best2} 0.925                     & \cellcolor{best2}27.23                     & \cellcolor{best1} 0.059                      \\ \hline
\end{tabular}

\caption{\textbf{Quantitative Results on NeRF Synthetic and Tanks \& Temple Datasets.} We compare methods with similar  parameter counts. 
We also include results with the full number of primitives for broader comparison.}
\label{tab:comparison-parameter}
\end{table*}

\begin{table}[ht]
\centering
\small
\begin{tabular}{c|cccc}
\hline
\multicolumn{1}{c|}{Method} & \multicolumn{1}{c}{\#Parameters} & {SSIM↑} & \multicolumn{1}{c}{PSNR↑} & \multicolumn{1}{c}{LPIPS↓} \\ \hline
3DGS                                          & 383.5K                     & 0.721                     & 20.66                     & 0.333                      \\
Mip-Splatting                                 & 383.5K                    & 0.727                     & \cellcolor{best2}20.94     & 0.344                      \\
Mini-Splatting                                & 339.8K                     & 0.711                     & 20.84                     & 0.357                      \\
Scafold-GS                                    & 360.1K                      & \cellcolor{best2}0.727    & 20.66                     & \cellcolor{best2}0.323    \\ \hline
\textbf{Ours}                                 & 343.5K                     & \cellcolor{best1}0.736    & \cellcolor{best1}21.19     & \cellcolor{best1}0.306    \\ \hline
\end{tabular}
\caption{\textbf{Quantitative Results on NeRF Synthetic dataset.} In the close-up view, our method better preserves boundaries, outperforming all other methods.}
\label{tab:comparison-closer}
\end{table}

We have further extended our evaluation to more complex objects and real scenes. Given the compact nature of our approach, we compare BG-Triangle with other methods using a low but comparable parameter count. All experiments are initialized with a coarse point cloud. The quantitative comparison is provided in Tab.~\ref{tab:comparison-parameter}. Under comparable parameter configurations, our method significantly outperforms purely Gaussian-based models, such as 3DGS \cite{kerbl3Dgaussians} and Mip-Splatting \cite{yu2024mip}. When compared with other compact methods like Mini-Splatting \cite{fang2024mini}, which adaptively controls primitive counts, and Scaffold-GS \cite{scaffoldgs}, which employs multi-layer perceptrons (MLPs) for compact representation, BG-Triangle still achieves comparable and often better performances. Our approach demonstrates significant advantages in terms of the LPIPS metric, suggesting that the images we generate are perceptually more similar and visually closer to the target. We have also included the results with their original versions for comparisons. The complete qualitative comparison is shown in Fig.~\ref{fig:comparison-tt}. In test views at distances similar to the training views, our method manages to preserves rendering details at a much higher fidelity.

\begin{figure*}[ht]
    \centering
    \includegraphics[width=\linewidth]{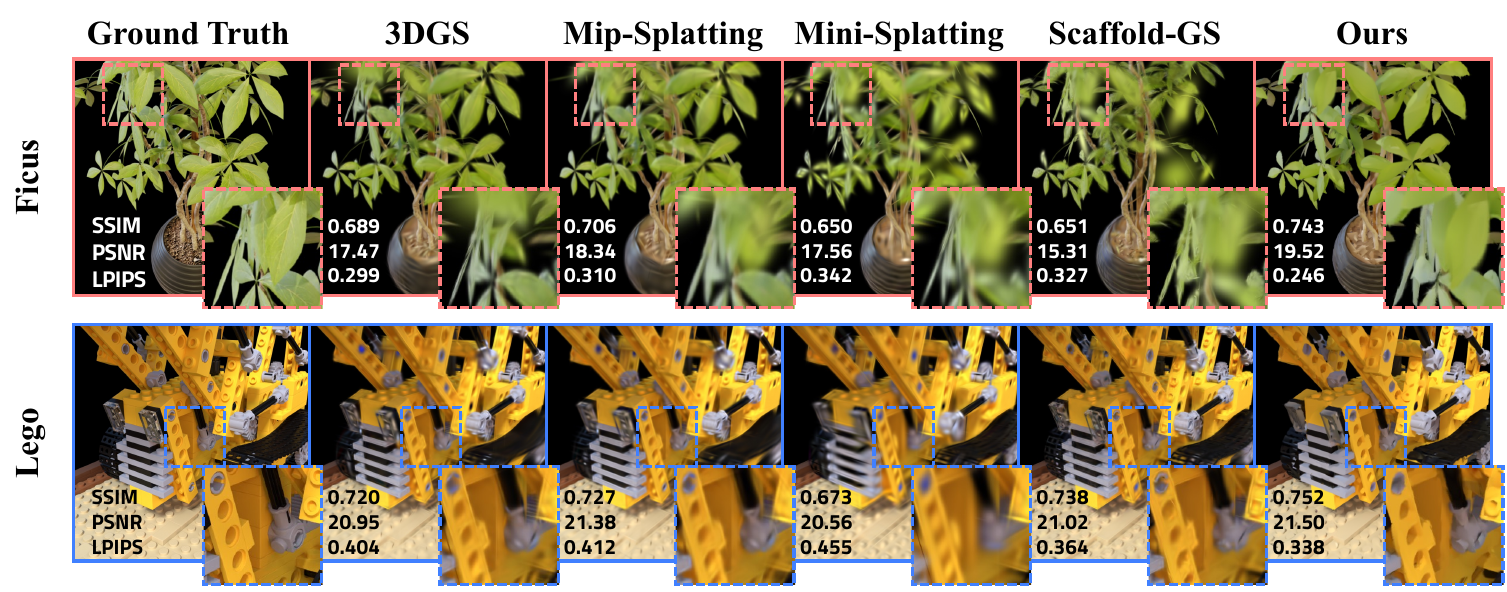}
    \caption{\textbf{Qualitative Results on the NeRF Synthetic Dataset.} We compare methods with a similar number of parameters in a close-up view and provide metrics for each view in this figure.}
    \label{fig:comparison-nerf}
\end{figure*}

To further evaluate the methods, we reduce the test view distance to 40\% of its original value while keeping the number of primitives low and the training views unchanged. The ground truth for the close-up views of the NeRF synthetic dataset is rendered using the released Blender projects. The quantitative results provided in Tab.~\ref{tab:comparison-closer} show that our method outperforms existing approaches under this setting. As shown in Fig.~\ref{fig:comparison-nerf}, even at closer test views, our method maintains clear object boundaries, whereas Gaussian-based methods tend to exhibit blurring or spiky artifacts. Notably, their approach and ours exhibit distinct optimization directions. Our method prioritizes maintaining clear boundaries, albeit at the cost of structural misalignment, while theirs tends to result in blurred boundaries.

\subsection{Ablation Study}

\begin{table}[ht]
\centering
\small
\begin{tabular}{c|cccc}
\hline
             & \#Parameters & SSIM↑ & PSNR↑     & LPIPS↓   \\ \hline
\textbf{Ours}   & 343.5K   & 0.937   & 29.16  & 0.050       \\
$\text{Ours}_{\text{low}}$ & 175.9K  & 0.931  & 28.31  & 0.059      \\
$\text{Ours}_{\text{high}}$  & 1.026M   & 0.943 & 29.67  & 0.047     \\
\hline
w/o SH   & 217.6K  & 0.929   & 27.54  & 0.058      \\
w/o B & 343.6K & 0.932 & 28.98  & 0.054       \\ 
\hline
\end{tabular}
\caption{\textbf{Ablation Study}. We compare the variants with different number of parameters and different modules. } \label{tab:ablate}
\end{table}

\begin{figure}[t]
    \centering
    \includegraphics[width=\linewidth]{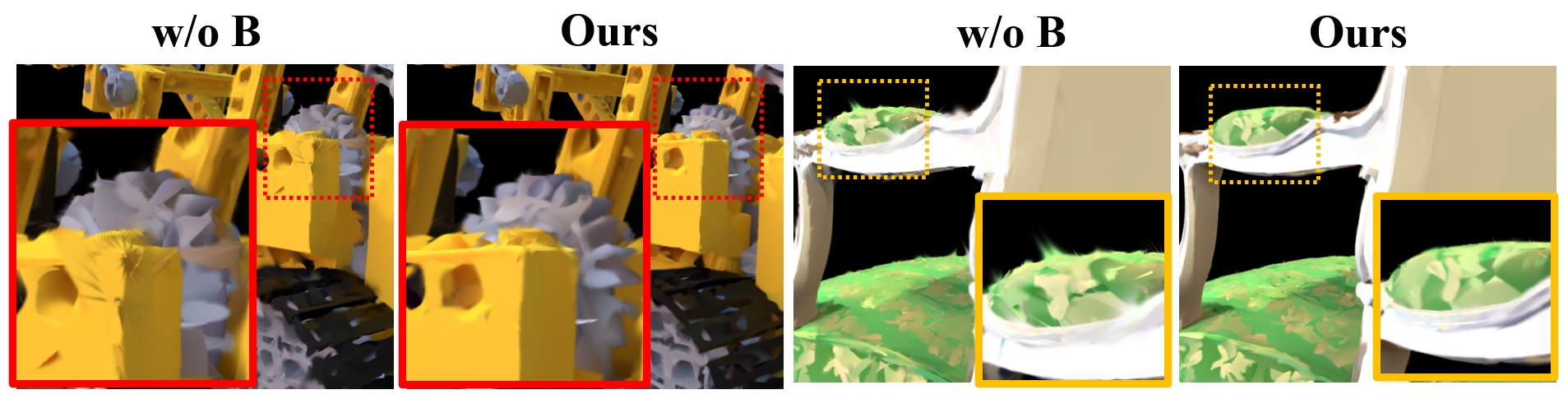}
    \caption{The rendering results of ``ours'' and ``w/o B'' under the zoom-in perspective. Our method is able to render clear boundaries even at close perspectives.}
    \label{fig:ablation}
\end{figure}

We ablate BG-Triangle's rendering quality under different settings on Nerf Synthetic Dataset, as shown in Tab.~\ref{tab:ablate}. We control the parameter size by setting the resolution of the attribute map. We experimented with two variants, $\text{Ours}_{\text{low}}$ and $\text{Ours}_{\text{high}}$, where the effective pixel count for each primitive's attribute map is 1 and 10, respectively.

Overall, the rendering quality increases with the parameter amount, but not in a strictly proportional manner. Although our method only allows a primitive to share a set of SH coefficients, this is necessary. Compared to a model without SH coefficients (w/o SH), there is a significant improvement in rendering quality. When training and testing without blending coefficients (w/o B), the results are very close to that of ours. However, upon closer inspection, noticeable artifacts appear, as shown in Fig.~\ref{fig:ablation}, suggesting that the Gaussians are overfitting to the camera views.

\vspace{2mm}
\textbf{Training Time and Memory Cost.} 
We report the cost on the Lego scene in Tab.~\ref{tab:training}. Under the default setting, the total training time is about 35 minutes on one NVIDIA RTX3090 GPU. During training, we use the CUDA rasterization software, which may not be very efficient and increases $t_{\text{BO}}$. The training memory usage remains unchanged due to PyTorch's memory management.

\begin{table}[h]
\centering
\begin{tabular}{cc|cccc|c}
\hline
\#P. & $r_{b}$ & $t_{\text{TR}}$ & $t_{\text{SG}}$ & $t_{\text{DAB}}$ & $t_{\text{BO}}$ & Mem. \\ \hline
6K        & 1e-3         & 14.69              & 1.37            & 13.42              & 35.33              & 6.8         \\
12K       & 1e-3         & 19.38              & 1.39            & 12.80              & 34.15              & 6.8         \\
6K        & 1e-2         & 14.73              & 1.34            & 26.91              & 50.90              & 6.8         \\
\hline
\end{tabular}
\caption{Training time of each stage per iteration (in ms) and memory cost (in GB) of our method. \#P stands for the number of primitives. $r_b$ is scaling factor for boundary width, $t_{\text{TR}}$ is tessellation and rasterization time, $t_{\text{SG}}$ is sub-primitive generation time, $t_{\text{DAB}}$ is the time for discontinuity-aware alpha blending, and $t_{\text{BO}}$ is backward and optimization time.}
\label{tab:training}
\end{table}

\section{Discussion and Conclusion}
The past five years have witnessed the rise of probabilistic representations such as NeRF, 3DGS, and various variations. As aforementioned, they correspond to primitives between two ends: completely deterministic and completely probabilistic. Our new primitive, BG-Triangle, lies somewhere in between and is slightly more ``geometric'' than 3DGS, as we impose more connectivity constraints between the primitives. BG-Triangle further leverages attribute sharing within primitives to achieve highly efficient expressiveness, enabling more effective scene representation and novel view synthesis, particularly when using much fewer number of primitives. 

Although visually comparable, BG-Triangle performs slightly worse than 3DGS in overall rendering quality, as measured by classic metrics such as PSNR and SSIM. However, we emphasize that these metrics do not fully capture important visual features like edges and boundaries, highlighting the need for more objective evaluation methods. A key advantage of BG-Triangle is its lightweight nature, requiring significantly fewer primitives and parameters.

%

\vspace{2mm}
\textbf{Acknowledgments.} {\raggedleft This work was supported by the Flanders AI Research program and NSFC programs (W2431046). We also acknowledge support from the Shanghai Frontiers Science Center of Human-centered Artificial Intelligence, MoE Key Lab of Intelligent Perception and Human-Machine Collaboration (ShanghaiTech University), and HPC Platform of ShanghaiTech University.}

{
    \small
    \bibliographystyle{ieeenat_fullname}
    \bibliography{main}
}


\clearpage
\setcounter{page}{1}
\maketitlesupplementary

\section{Implementation Details}

\paragraph{Code Release}

We built upon the 3DGS \cite{kerbl3Dgaussians} code base to add discontinuity-aware rendering and corresponding acceleration algorithms in the CUDA rendering code. We used PyTorch \cite{paszke2019pytorch} to implement the proposed rendering pipeline. Our code will be made publicly available to the community once the paper is accepted.

\paragraph{Point Cloud Initialization}

Our method can be initialized using either a mesh or a coarse point cloud. Generally, point clouds are easier to obtain compared to meshes. A coarse point cloud can be generated from MVS \cite{MVS} algorithms, 3D scanners, or even 3DGS. In our experiments, for each scene, the point cloud we used was derived from the reconstruction result of the 3000th iteration of 3DGS. We randomly sampled 10,000 points from this result to form the initial point cloud. On each point in the initial point cloud, we placed equilateral triangle primitives with different orientations but the same size as the initialization result.

\begin{figure}[ht]
    \centering
    \includegraphics[width=\linewidth]{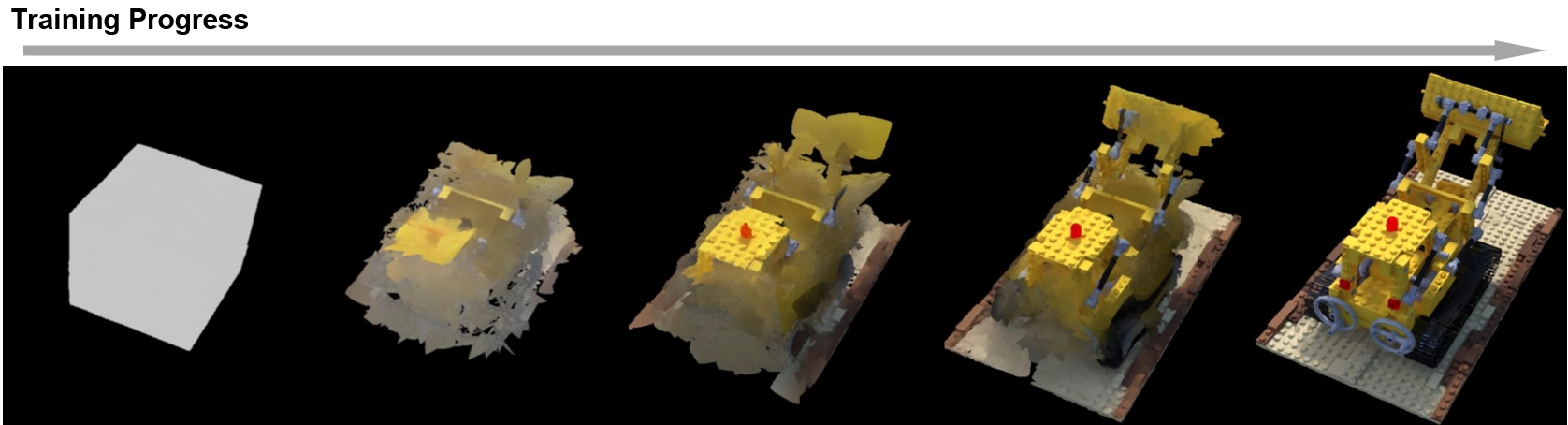}
    \caption{\textbf{The visualization of cube initialization} of the training progress on the Lego scene from the NeRF-Synthetic dataset.}
    \label{fig:cube_init}
\end{figure}

\paragraph{Cube Initialization}
BG-Triangle can reconstruct scenes initialized from a cube (Fig.~\ref{fig:cube_init}). 
During optimization, BG-Triangles dynamically `grow' and `contract' to conform to the shape of the 3D scene.

Optimization with cube initialization requires more iterations to match the reconstruction quality of point-cloud initialization and is more sensitive to hyperparameters (e.g., learning rates). Since point clouds are easily generated from multi-view images during calibration, we use point cloud initialization in our experiments.

\paragraph{Multi-Resolution Attribute maps.}
In our implementation, the attribute map of each primitive is a multi-channel grid shaped as an isosceles right triangle, with the pixel resolution of the legs being 3, except the SH attribute map whose resolution is 1. The number of channels in the attribute map corresponds to the original dimension of the respective attribute.

\paragraph{Splitting and Pruning.} 
Splitting and pruning are essential for Bézier Gaussian Triangles (BG-Triangles) to effectively represent a scene. The splitting and pruning strategy is performed every 300 iterations and is disabled after the 15,000th iteration to stabilize the optimization process.

For splitting, we follow two pre-defined criteria: position gradient amplitude and edge prior. We compute the position gradient for each BG-Triangle. If the average position gradient of a BG-Triangle exceeds \(\tau_g = 0.0018\), the region is considered under-reconstructed. Using the index map \(\mathbf{I}_{id}\), we map the edge gradients of the image to the corresponding BG-Triangle and accumulate their values. If this accumulated average exceeds \(\tau_b = 13.0\), it indicates that the BG-Triangle may be crossing a boundary. In either case, we subdivide the BG-Triangle into four smaller BG-Triangles by splitting it at the midpoints of its edges.

For pruning, we follow three pre-defined criteria: visibility, area, and aspect ratio of the primitives. Using the index map \(\mathbf{I}_{id}\), the visibility count of each BG-Triangle is tracked. When the average visibility count ratio falls below \(\tau_v = 0.08\), the BG-Triangle is considered occluded. To further evaluate visibility, we maintain a visibility texture for each BG-Triangle. Using the coordinate map \(\mathbf{I}_{uv}\), the UV coordinates are mapped to the visibility texture, and the proportion of visible area is calculated. If the visibility ratio falls below \(\tau_r = 0.4\), the BG-Triangle is also considered occluded. Additionally, Bézier triangles with an area smaller than \(\tau_a = 3 \times 10^{-4}\) or with an excessively elongated shape (aspect ratio above \(\tau_s = 10\)) are considered insignificant or noisy. The area is estimated by approximating the BG-Triangle as four planar triangles and summing their areas, while the aspect ratio is determined by computing the aspect ratio of the bounding box of BG-Triangle. In any of these cases, the corresponding BG-Triangles are pruned.

\section{Additional Results}
\paragraph{Reconstruction Quality}

We compare the reconstruction quality using Chamfer Distance on the NeRF-Synthetic dataset between our method and 2DGS \cite{Huang2DGS2024} under its default settings for bounded scene, with the downsampling density set to 0.2. As shown in Table \ref{tab:chamfer}, our method achieves comparable results to 2DGS.

\begin{table}[h]
\centering
\begin{tabular}{c|cccccccc}
\hline
     & Chair           & Drums           & Ficus           & Hotdog          \\ \hline
2DGS & \textbf{0.0645} & 0.0675          & 0.0667          & 0.0693          \\ \hline
Ours & 0.0722          & \textbf{0.0592} & \textbf{0.0535} & \textbf{0.0642} \\ 
\hline
     & Lego            & Material        & Mic             & Ship            \\ \hline
2DGS & \textbf{0.0581} & \textbf{0.0540} & 0.0629          & \textbf{0.0735} \\ \hline
Ours & 0.0597          & 0.0561          & \textbf{0.0628} & 0.0786          \\ \hline
\end{tabular}
\caption{Reconstruction quality comparisons on NeRF-Synthetic Dataset.}
\label{tab:chamfer}
\end{table}

\paragraph{Applications}

\begin{figure}[ht]
    \centering
    \includegraphics[width=\linewidth]{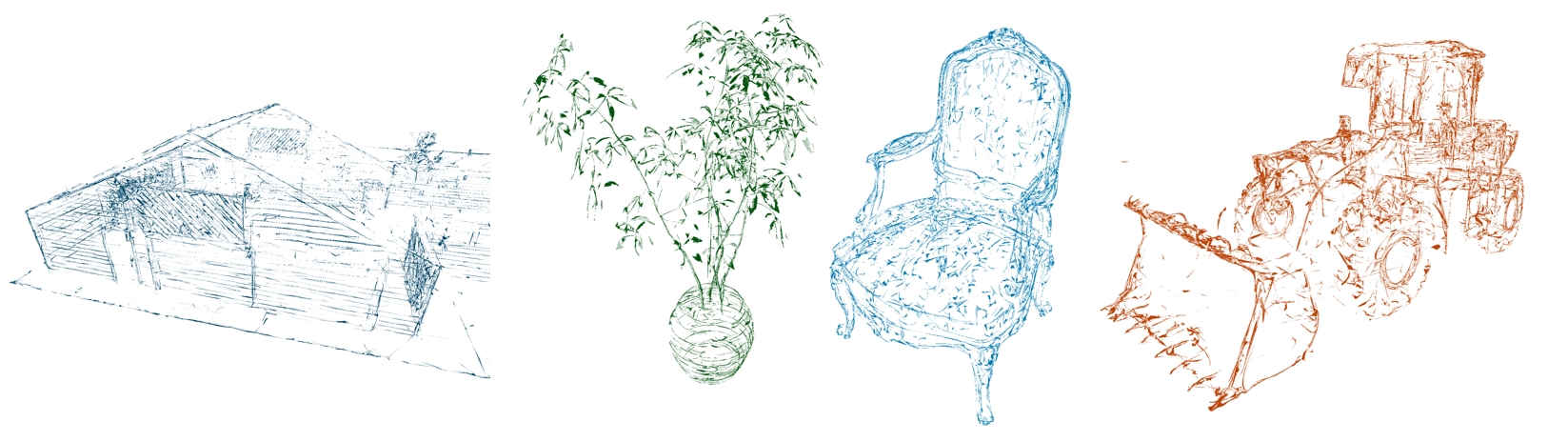}
    \caption{3D line strokes extracted from dense boundary primitives reveal rich contour information and semantic features of the scene.}
    \label{fig:strokes}
\end{figure}

The 3D vector representation reconstructed by BG-Triangle from multi-view images shows a clustering effect at the geometric and texture boundaries. In other words, there is a relatively dense distribution of primitives at the boundaries. By leveraging this distribution characteristic, we can extract the 3D line strokes at the boundaries by filtering out tessellated primitives based on their face area and removing those with areas that are too large. As shown in the Fig.~\ref{fig:strokes}, the 3D line strokes allow us to easily distinguish the content of the scene, indicating that these 3D line strokes contain rich contour information and exhibit certain semantic features.

\section{Backward Derivative}

\begin{figure*}[htp]
   \centering
   \includegraphics[width=\linewidth]{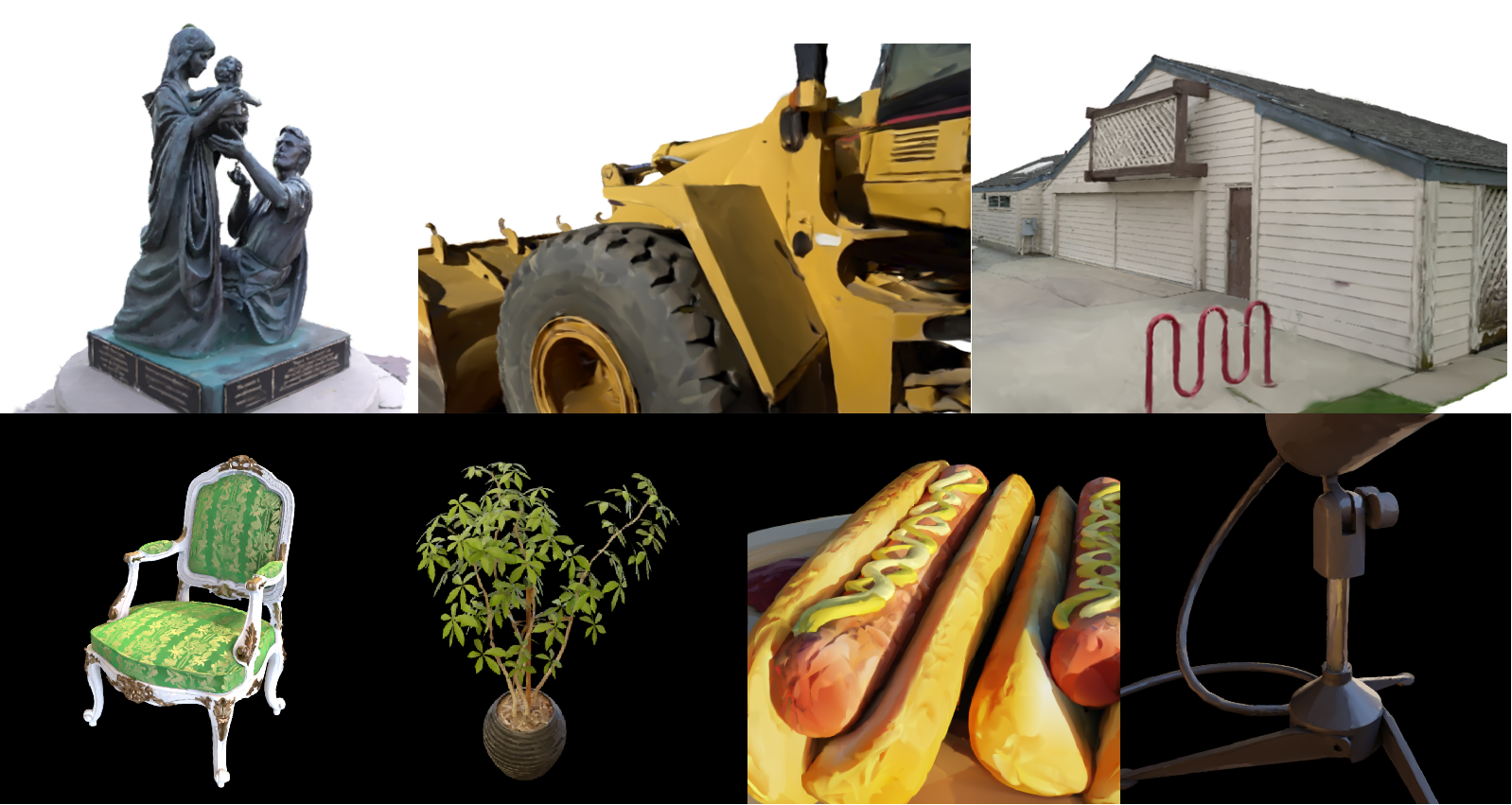}
   \caption{\textbf{Additional results of our BG-Triangle.}}
   \label{fig:gallery}
\end{figure*}

We follow the notations defined in Sec. 3.

\paragraph{Derivative of Gradient in Discontinuity-aware Alpha Blending.}

The alpha value with blending coefficient is:
\begin{equation}
\alpha(\mathbf{q}) = o \cdot w(\mathbf{q}) \cdot \exp\left(-\frac{1}{2}(\mathbf{q} - \mu)^{\top}\Sigma^{-1}(\mathbf{q} - \mu)\right)
\end{equation}
where $\mu$ is the 2D position of the Gaussian $\mathcal{G}$, $\Sigma$ is the 2D covariance matrix, $o$ is a constant scalar. In alpha blending, the final color is computed as:
\begin{equation}
C = \sum_{i = 0}^{n - 1}T_{i}\alpha_{i}c_{i} + T_{n}c_{\text{bg}}
\end{equation}
where $c_{\text{bg}}$ is the background color, $c_{i}$ is the color associated with each Gaussian $\mathcal{G}_{i}$, $T_{i}$ is the transparent term, defined as:
\begin{equation}
T_{i} = \prod_{j = 0}^{i - 1}(1 - \alpha_{j}), \quad T_{0} = 1
\end{equation}

The gradient of the loss $\ell$ with respect to color $c_{i}$ is given by:
\begin{equation}
\frac{\partial\ell}{\partial{c_{i}}} = \frac{\partial\ell}{\partial{C}} \frac{\partial{C}}{\partial{c_{i}}} = \frac{\partial\ell}{\partial{C}} T_{i}\alpha_{i}.
\end{equation}

The gradient of $\ell$ with respect to alpha $\alpha_{i}$ is:
\begin{equation}
\begin{aligned}
\frac{\partial\ell}{\partial{\alpha_{i}}} &= \frac{\partial\ell}{\partial{C}}\frac{\partial{C}}{\partial{\alpha_{i}}} \\
&= \frac{\partial\ell}{\partial{C}} \left[\left(\sum_{j = 0}^{n - 1}\frac{\partial{T_{j}}}{\partial{\alpha_{i}}}\alpha_{j}c_{j} + T_{j}\frac{\partial{\alpha_{j}}}{\partial{\alpha_{i}}}c_{j}\right) + \frac{\partial{T_{n}}}{\partial{\alpha_{i}}}c_{\text{bg}}\right] \\
&= \frac{\partial\ell}{\partial{C}} \left[T_{i}c_{i} - \frac{1}{1 - \alpha_{i}}\left(\sum_{j = i + 1}^{n - 1}T_{j}\alpha_{j}c_{j} + T_{n}c_{\text{bg}}\right)\right].
\end{aligned}
\end{equation}

The gradient of $\ell$ with respect to $w$ is:
\begin{equation}
\frac{\partial\ell}{\partial{w}} = \frac{\partial\ell}{\partial{\alpha}} \frac{\partial{\alpha}}{\partial{w}} = \frac{\partial\ell}{\partial{\alpha}}\frac{\alpha}{w}.
\end{equation}

Boundary pixels, denoted as $\mathcal{B}$, are derived from the index map $\mathbf{I}_{\text{id}}$, which is ultimately determined by the control points of Bézier triangles. To account for these boundary pixels, we calculate the gradient of $\ell$ with respect to a boundary pixel $\mathbf{b}\in\mathcal{B}$:
\begin{equation}
\begin{aligned}
\frac{\partial\ell}{\partial\mathbf{b}} &= \frac{\partial\ell}{\partial{w}}\frac{\partial{w}}{\partial\mathbf{b}} \\
&= \begin{cases}
0 & \text{if } \mathbf{I}_{id}(\mathbf{b}) <> e,\\
\dfrac{\partial\ell}{\partial{w}}\gamma'(\|\mathbf{q} - \mathbf{b}\|; \sigma)\dfrac{\mathbf{q} - \mathbf{b}}{\|\mathbf{q} - \mathbf{b}\|} & \text{if } \mathbf{I}_{id}(\mathbf{b}) = \mathbf{I}_{id}(\mathbf{q}) \\
-\dfrac{\partial\ell}{\partial{w}}\gamma'(\|\mathbf{q} - \mathbf{b}\|; \sigma)\dfrac{\mathbf{q} - \mathbf{b}}{\|\mathbf{q} - \mathbf{b}\|} & \text{otherwise}
\end{cases}
\end{aligned}
\end{equation}
where
\begin{equation}
\begin{aligned}
\gamma'(\|\mathbf{q} - \mathbf{b}\|; \sigma) &= \frac{\ln2}{\sigma}2^{\frac{\|\mathbf{q} - \mathbf{b}\|}{\sigma} - 1} \\
&= \frac{\ln2}{\sigma}\gamma(\|\mathbf{q} - \mathbf{b}\|; \sigma).
\end{aligned}
\end{equation}

Other factors related to the attributes of the Gaussian $\mathcal{G}$ can be handled automatically by the 3DGS framework, and are therefore omitted here for brevity.

\paragraph{Derivative of Gradient in Sub-primitive Generation.}

Both the 3D position \( \mathbf{S}_{\mathbf{q}} \) and the diffuse color \( \mathbf{c}_{\mathbf{q}} \) of the Gaussian \( \mathcal{G} \) can be interpolated barycentrically by the control points. Taking the 3D position \( \mathbf{S}_{\mathbf{q}} \) as an example, the gradient of the loss \( \ell \) with respect to a control point \( \mathbf{p}_{i, j, k}(\mathbf{I}_{id}(\mathbf{q})) \) is given by:

\begin{equation}
\begin{aligned}
\frac{\partial\ell}{\partial{\mathbf{p}_{i, j, k}}(\mathbf{I}_{id}(\mathbf{q}))} &= \frac{\partial\ell}{\partial{\mathbf{S}_{\mathbf{q}}}} \frac{\partial\mathbf{S}_{\mathbf{q}}}{\partial{\mathbf{p}_{i, j, k}}(\mathbf{I}_{id}(\mathbf{q}))} \\
&= \frac{\partial\ell}{\partial{\mathbf{S}_{\mathbf{q}}}} B_{i, j, k}^{n}(\mathbf{I}_{uv}(\mathbf{q})).
\end{aligned}
\end{equation}
where $B_{i, j, k}^{n}$ is the Bernstein polynomial of degree \(n\) defined on the barycentric coordinate system.

\section{Result Gallery}
We provide additional results of our BG-Triangle in Fig. \ref{fig:gallery}.

\section{Comparison Setup}

We first set an upper control on the number of primitives for different methods based on a predefined parameter scale. Starting with sparse point clouds, we optimize and proceed to the splitting phase. During this phase, we check the total number of primitives after splitting. If the total number remains within the control, we keep the splitting results. If it exceeds the control, we randomly select newly added primitives, ensuring the total number adheres to the control.

In addition, pruning operations for each method remain enabled to dynamically adjust the primitive distribution effectively. When pruning, we disable pruning large Gaussian primitives to ensure that different methods can fit the overall structure of the object. This approach balances dynamic adjustments with fitting precision.

\section{Limitations}

Similar to the vectorization process for 2D images, which often entails a loss of image quality, the representations reconstructed and rendered using our method exhibit a quality that remains slightly below that achieved by state-of-the-art novel view synthesis techniques. At the same time, our approach, which generates only a single layer of sub-primitives, cannot handle materials with opacity.



\end{document}